\documentclass[letterpaper]{article} 
\usepackage{aaai25}  
\usepackage{times}  
\usepackage{helvet}  
\usepackage{courier}  
\usepackage[hyphens]{url}  
\usepackage{graphicx} 
\usepackage{natbib}  
\usepackage{subcaption}
\usepackage{amsmath}
\usepackage{caption} 
\usepackage{algorithm}
\usepackage{algorithmic}
\usepackage{multirow}
\usepackage{newfloat}
\usepackage{listings}
\DeclareCaptionStyle{ruled}{labelfont=normalfont,labelsep=colon,strut=off} 
\floatstyle{ruled}
\newfloat{listing}{tb}{lst}{}
\floatname{listing}{Listing}
\title{mRNA2vec: mRNA Embedding with Language Model in the 5$'$ UTR-CDS for mRNA Design}
\author{
    Honggen Zhang\textsuperscript{\rm 1},
    Xiangrui Gao\textsuperscript{\rm 2},
    June Zhang\textsuperscript{\rm 1},
    Lipeng Lai\textsuperscript{\rm 2}\thanks{Corresponding Author}
}
\affiliations{
    \textsuperscript{\rm 1}University of Hawaii at Manoa\\
    \textsuperscript{\rm 2}XtalPi Inc\\

    \{honggen,zjz\}@hawaii.edu, \{xiangrui.gao, lipeng.lai\}@xtalpi.com
}

\usepackage{bibentry}

\begin{document}

\maketitle

\begin{abstract}
Messenger RNA (mRNA)-based vaccines are accelerating the discovery of new drugs and revolutionizing the pharmaceutical industry. However, selecting particular mRNA sequences for vaccines and therapeutics from extensive mRNA libraries is costly. Effective mRNA therapeutics require carefully designed sequences with optimized expression levels and stability. This paper proposes a novel contextual language model (LM)-based embedding method: mRNA2vec. In contrast to existing mRNA embedding approaches, our method is based on the self-supervised teacher-student learning framework of data2vec. We jointly use the 5$'$ untranslated region (UTR) and coding sequence (CDS) region as the input sequences. We adapt our LM-based approach specifically to mRNA by 1) considering the importance of location on the mRNA sequence with probabilistic masking, 2) using Minimum Free Energy (MFE) prediction and Secondary Structure (SS) classification as additional pretext tasks. mRNA2vec demonstrates significant improvements in translation efficiency (TE) and expression level (EL) prediction tasks in UTR compared to SOTA methods such as UTR-LM. It also gives a competitive performance in mRNA stability and protein production level tasks in CDS such as CodonBERT.
The code is available in GitHub\footnote{https://github.com/honggen-zhang/mRNA2vec}
\end{abstract}

%

\section{Introduction}
Messenger RNA (mRNA)-based gene therapeutics and vaccines have emerged as pivotal tools for developing a new class of drugs, as they do not present the safety concerns associated with genomic integration \cite{to2021overview}. Recently, mRNA vaccines have been utilized in treating several diseases, including COVID-19 \cite{jackson2020mrna} and cancer \cite{zeraati2017cancer}. mRNA therapeutics facilitate the production of missing or defective proteins, providing essential supplements to cure genetic diseases \cite{kwon2018emergence}. To advance mRNA-based treatments further, understanding and designing the appropriate mRNA sequences is crucial. Factors such as expression level and stability significantly impact the practical application of mRNA-based therapies.

Researchers have primarily focused on the untranslated region (UTR) and the coding sequence (CDS) region to understand the mechanisms of mRNA in protein translation \citet{cao2021high,li2023codonbert}. Studies have demonstrated that UTR can regulate mRNA translation \cite{akiyama2022informative}, with extensive exploration of the 5$'$ UTR, which precedes the coding sequences \cite{chu20245, wieder2024differences}. The CDS region contains codons that are translated into amino acids, so codon optimization directly impacts mRNA translation.

Technological advancements using machine learning (particularly language models) with biology have enabled new methods to decode mRNA sequences \cite{diez2022icodon, cao2021high}. Ribosomes read sequences from left to right, similar to text sequences, suggesting that mRNA sequences could be modeled using language models and self-supervised learning techniques. Mapping mRNA to a high-dimensional representation space has the potential to enhance our understanding of its functions. However, significant differences exist between mRNA and natural text. For example, mRNA is composed of only four nucleotides, and there are only 64 codons. Therefore it is important to account for these differences when adapting language models for mRNA embedding.


In this paper, we propose an mRNA embedding method: mRNA2vec. Unlike previous research that analyzed the 5$'$ UTR and CDS separately, we treat them as a single sequence input. The junction between the tail of the 5$'$ UTR and the head of the CDS plays a significant role in regulating mRNA translation  \cite{nieuwkoop2023revealing}. Utilizing information from both sides will enhance mRNA sequence representation. Instead of employing a single masked token prediction-based language model, we perform our model on a contextual level, such as data2vec\cite{baevski2022data2vec}, which considers both masked and unmasked sequences, providing richer sequence information for downstream tasks as it reflects the unmasked nature of sequence prediction.

In addition to the unifying 5$'$ UTR and CDS and contextual representation learning, we fully utilize the properties of the mRNA modality: 1) We specify a probabilistic hard-masking strategy that focuses more on the tail region of the 5$'$ UTR and head of the CDS, increasing the likelihood of these regions being masked. 2) We incorporate additional modalities, such as Minimum Free Energy (MFE) and Secondary Structure (SS), to design better pretraining tasks to enhance the process.

To evaluate mRNA2vec, we consider translation expression tasks on both 5$'$ UTR and CDS datasets. For the 5$'$ UTR, we assess Translation Efficiency (TE) and Expression Level (EL) as downstream tasks. For the CDS, we focus on mRNA stability and protein production level as the benchmarks. Our experiments demonstrate significant improvements over state-of-the-art results for the TE and EL tasks and comparable results for mRNA stability and protein production level.

Moreover, we discuss the reasons mRNA2vec is successful in translation-related tasks such as TE. Designing the pretext task, selecting specific regions of the input sequence, choosing better hidden states, and designing a complex downstream task regressor can help to improve experimental outcomes. 
In summary, our main contributions are:
\begin{enumerate}
    \item We propose a novel embedding method, mRNA2vec, a unified method that combines the 5$'$ UTR and CDS regions as the input sequence. It is also the first application of the data2vec pretraining method to mRNA sequences. Experimental results show contextual learning will better learn mRNA sequence representation than the single mask token prediction.
    \item We propose a probabilistic hard-masking strategy based on mRNA properties for comparing the representation of the masked region from the student model and the output from the teacher model. Experimental results show that hard masking can help decrease loss and improve the alignment between representation and MFE.
    \item We designed two additional pretext tasks based on MFE and SS, different from UTR-LM \cite{chu20245}. Experimental results show that our utilization of MFE and SS can consistently improve the embedding results.
    \item We fine-tune our model on downstream tasks for both 5$'$ UTR and CDS datasets. It shows that mRNA2vec is superior for the tasks compared to the SOTA benchmarks such as UTR-LM and CodonBERT. 
\end{enumerate}

\begin{figure*}[h!]
     \centering
    \includegraphics[width=0.80\textwidth]{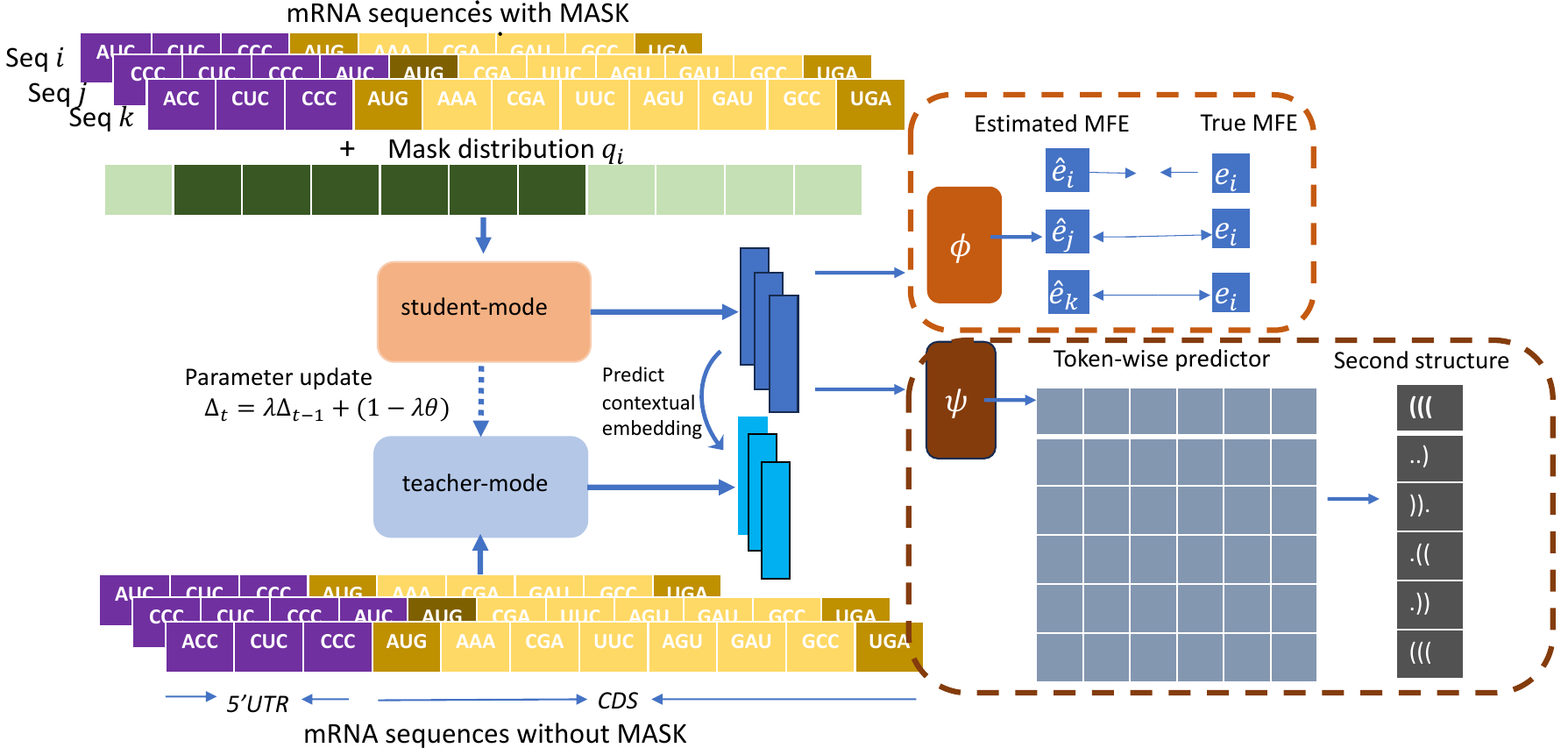}
    \caption{The diagram for the pretraining. The data2vec teacher model parameter $\Delta$ will be updated by the current student model parameter $\theta$ using EMA. The input sequence will be masked based on the hard-mask distribution $q$. 
    The $\phi$ and $\psi$ are the neural networks used for MFE regression and SS classification.}
    \label{fig:diagram1}
\end{figure*}

\section{Related Work}
\subsection{mRNA Sequence-based Translation Prediction}

The optimization of mRNA sequences to enhance expression and stability has become increasingly viable, thanks to the vast amount of measured and synthesized sequences available \cite{2023Deep}. Data-driven models have emerged as effective for predicting synonymous sequence expression, as they can detect highly nonlinear correlations between sequences and their functional outcomes. For instance, \citet{leppek2022combinatorial} introduced the RNA sequencing-based platform PERSIST-seq, which systematically predicts mRNA stability and ribosome load. Similarly, \citet{diez2022icodon} developed an evolutionary algorithm for codon optimization to address the challenge posed by the enormous number of possible synonymous sequences that encode the same peptide. In other studies, \citet{cao2021high} and \citet{nieuwkoop2023revealing} employed machine learning approaches, specifically random forests, to model the 5’ UTR and CDS regions, predicting translation efficiency and protein production levels, respectively. 
Moreover, the length of the 5’ UTR sequence itself can be a direct predictor of mRNA translation regulation \cite{wieder2024differences}.  
Recently, the success of BERT-based \cite{devlin2018bert} pre-trained language models across various NLP tasks has inspired similar approaches for translation prediction tasks, such as RNA-FM \cite{chen2022interpretable} and RNABERT \cite{akiyama2022informative}. These BERT-based methods primarily rely on masked token prediction during pretraining. However, given the specific properties and translation prediction tasks associated with mRNA, we prioritize learning a comprehensive sequence representation over token-level predictions. Contextual learning-based pre-trained language model architectures, such as data2vec \cite{baevski2022data2vec}, remain largely unexplored in this domain.

\subsection{mRNA Sequence Representation Learning}

Self-supervised learning and language models have recently emerged as powerful tools for extracting sequence features from large, unlabeled datasets \cite{devlin2018bert,raffel2020exploring}. DNA2vec \cite{ng2017dna2vec}, inspired by word2vec \cite{mikolov2013efficient}, introduced a method for learning distributed representations of k-mers. 
For base-by-base resolution embedding, RNA-BERT \cite{akiyama2022informative} adapts the pretraining algorithm BERT to non-coding RNA (ncRNA), embedding the four RNA bases while combining the Masked Language Model task with secondary structure alignment. Beyond mere representation, recent works try to understand the expression of mRNA by applying these representations. For example, \citet{2023Deep} developed a codon optimization language model for the Enterobacterales data set by embedding codons, while CondoBERT \cite{li2023codonbert} further extended this approach by incorporating more pretraining data sets.
By predicting masked tokens at the codon level, CondoBERT successfully obtains codon embeddings.
Although the CDS region provides substantial information for predicting mRNA translation, the 5’ UTR, as a regulatory sequence, also plays a critical role in learning translation efficiency \cite{nieuwkoop2023revealing,2021Systematic,2018Human}. Recently, UTR-LM \cite{chu20245} was proposed to specifically focus on the 5’ UTR sequence to train a language model and  RiNALMo \cite{penic2024rinalmo} provides the largest pre-trained model (with 650M parameters) on 36M non-coding RNA sequences.

\section{Method:mRNA2vec}
In this section, we introduce mRNA2vec, a novel mRNA language model that involves concatenating the 5$'$ UTR and CDS sequences to form a single input sequence. Unlike traditional models such as BERT \cite{devlin2018bert} or T5 \cite{raffel2020exploring}, which use a masked sequence as input, mRNA2vec adopts a contextual target-based pretraining model inspired by data2vec. This approach allows the model to access both the unmasked and masked sequences simultaneously.

We employ a probabilistic hard masking strategy in which specific regions of the sequence, particularly biologically important areas, have a higher likelihood of being masked. This ensures that these critical regions are taught more effectively during training. Additionally, domain-specific information, such as Minimum Free Energy (MFE) and Secondary Structure (SS), is incorporated through the definition of auxiliary pretraining tasks. This integration strengthens the model’s ability to capture a representation across the entire sequence, enhancing its utility for translation-related downstream tasks on the both 5$'$ UTR and CDS regions.

\subsection{mRNA Pre-training}

Pre-training has been very successful in advancing various domains, such as natural language understanding \cite{devlin2018bert,rao2020transformer}, image feature learning \cite{chen2020simple,he2020momentum}, and maintenance of tabular data sets\cite{huang2023searching,rubachev2022revisiting}. With a properly designed pretext task in the pretraining process, we will obtain the generalized representation for the data input. mRNA as the sequential dataset, we might discover some hidden information from the large data pretraining similar to the language. 

\subsubsection{pretraining Architecture}

The pre-training architecture of mRNA2vec is illustrated in Figure \ref{fig:diagram1}. Instead of focusing solely on the straightforward task of predicting single masked tokens, we employ the data2vec pretraining architecture to learn contextual targets from the unmasked sequence. This approach enables the model to capture the entire sequence information rather than just the masked parts, leading to a more comprehensive representation for downstream tasks involving unmasked sequences.

For tokenization, we utilize specialized tokens tailored to mRNA sequences. Since mRNA is composed of only four types of nucleotides ${\text{A},\text{U},\text{G},\text{C}}$, and is read by the ribosome in groups of three nucleotides, we design a tokenizer consisting of 64 tokens, each representing a specific sequence of three nucleotides (e.g., ${\text{AAA},\text{AAU},\text{AAG},\ldots}$). The tokenization applies on the UTR and CDS separately; then we concatenate them together. 

Instead of uniformly selecting masked tokens, as is common in NLP and CV tasks, we adopt a location-dependent probabilistic hard-masking strategy. Research has shown that regions such as the upstream CDS and the start of the CDS are crucial for mRNA translation \cite{nieuwkoop2023revealing,leppek2022combinatorial}. Therefore, tokens $X_i$ in these important regions are masked with a higher probability, as defined by:

\begin{equation}\label{eq:q}\small
p(X_i = \text{masked}) = q_i =
\begin{cases}
\frac{\alpha}{h}, & \mbox{if } i \in [k,k+h] \\
\frac{1-\alpha}{\ell-h}, & \mbox{otherwise},
\end{cases}
\end{equation}

where $\ell$ is the length of the input sequence, $h$ is the length of the continuous important region, and $\alpha$ adjusts the proportion of masked tokens selected from these regions.

By integrating the data2vec architecture with this specialized hard-masking strategy, mRNA2vec aims to reconstruct the entire sequence information based on the masked, crucial UTR-CDS regions. The loss function of the student model to predict the masked token given unmasked context from the teacher model for data2vec is defined as:

\begin{equation}\small
\mathcal{L} = ||f_{t}(x)-y||^2_{2},
\end{equation}

where $||\cdot||^2_2$ denotes the squared L2 norm. Here, $f_t$ represents the prediction of student encoder learning from the masked sequence, while $y$ is the representation obtained from the unmasked sequence, produced by a teacher model. The parameters of the teacher model are updated based on the student model through an exponentially moving average (EMA) \cite{he2020momentum}. Multiple hidden states are averaged to forms the contextual representation $y$ of the unmasked sequence. 

\subsubsection{Auxiliary Pretext Tasks with MFE and SS}
In addition to the primary sequence reconstruction pretext task, we incorporate auxiliary tasks involving Minimum Free Energy (MFE) and secondary structure (SS), which are known to correlate with mRNA expression levels \cite{akiyama2022informative,chu20245}. 

mRNA secondary structure consists of unpaired bases, represented by dots, which indicate single-stranded regions, and paired bases, represented by matching brackets, which form stem structures. For example, the SS of the sequence GGGAAACCC is represented as ``(((...)))''. MFE calculates base pair probabilities within thermodynamic models, representing the energy required to form the most stable secondary structure. To enhance mRNA2vec's ability to represent mRNA sequences for translation-related tasks, we introduce two additional pretext tasks:

\begin{enumerate}
    \item \textbf{MFE Regression}: We apply the InfoNCE loss \cite{oord2018representation} to capture the distance $d$ between the estimated MFE and the true MFE:

    \begin{equation}\small
    \mathcal{L}_{mfe} = -\log \sum_i \frac{\exp(-d(\widehat{e}_i,e_i))}{\sum_j \exp(-d(\widehat{e}_j,e_i))}
    \end{equation}

    Here, $\widehat{e}_i$ is the output scalar of the regressor $\phi(\cdot)$, applied on top of the pre-trained model. In this context, $\widehat{e}_i$ and $e_i$ form a positive pair, while $\widehat{e}_j$ and $e_i$ form a negative pair. The goal is to minimize the distance $d(\widehat{e}_i,e_i)$ while maximizing $d(\widehat{e}_i,e_j)$, learning to rank the predictions rather than merely minimizing $d(\widehat{e}_i,e_i)$ to zero. We use the mean token embedding from the last layer as the sequence representation, which serves as the input to $\phi(\cdot)$.

    \item \textbf{Secondary Structure Classification}: We treat SS prediction as a token-wise classification problem. Similar to nucleotide tokenization, we tokenize consecutive groups of three ``dot-bracket'' characters, resulting in a 27-category ($3\times 3\times 3$) classification problem:

    \begin{equation}\small
    \mathcal{L}_{ss} = \sum_i c_i\log(p_i)
    \end{equation}

    In this equation, \(p_i\) represents the softmax output of the classifier \(\psi(\cdot)\), applied on top of the pre-trained model, while \(c_i\) is a 27-dimensional one-hot vector representing the true label. The same token representation used as input to \(\phi(\cdot)\) is used as input to \(\psi(\cdot)\).
\end{enumerate}

Thus, the overall loss function for our pretraining model, mRNA2vec, is defined as:

\begin{equation}\small\label{eq:loss}
\mathcal{L}_{mRNA2vec} =  \mathcal{L}+\beta_{1}\mathcal{L}_{mfe}+\beta_{2}\mathcal{L}_{ss}
\end{equation}

Here, \(\beta_1\) and \(\beta_2\) are hyperparameters that regulate the contribution of MFE and SS to the language model. In this work, we set \(\beta_1 = 0.01\) and \(\beta_2 = 0.001\). 

\subsection{mRNA Translation Downstream Tasks}

After obtaining the pre-trained embedding model, we explore the representation applied to the downstream task related to mRNA translation. We can evaluate our model on both UTR sequence optimization and CDS region codon optimization. 

\begin{enumerate}
    \item \textbf{Evaluate mRNA2vec in 5$'$ UTR}:
    Understanding the mechanism of 5$'$ UTR 
influences in mRNA translation is crucial. In the 5$'$ UTR, we have two tasks to be predicted: mRNA Expression Level (EL) and  Translation Efficiency (TE)\cite{chu20245}. EL is measured in RNA-seq RPKM and TE is calculated as the ratio of Ribo-seq PRKM to RNA-seq RPKM. Ribo-seq is the ribosomal footprints observed on a given mRNA of interest. RNA-seq is the relative abundance of mRNA transcript in the cell. PRKM represents Reads Per Kilobase of transcript per Million mapped reads \cite{cao2021high}.
    
    \item \textbf{Evaluate mRNA2vec in CDS}: The codon optimization of the CDS region will directly affect the translation of mRNA sequence. Codon optimization refers to choosing the optimal amino acid codes for increasing protein expression. However, some proteins might need a slower translation rate to fold properly and maintain protein stability \cite{to2021overview}. Thus, in this region, Protein Production Level and mRNA Stability are used as two downstream tasks to measure the codon optimization of the CDS sequence.

\end{enumerate}

\section{Experiments}

\begin{figure*}[t]
     \centering
     \begin{subfigure}[t]{0.25\textwidth}
         \centering
         \includegraphics[width=\textwidth]{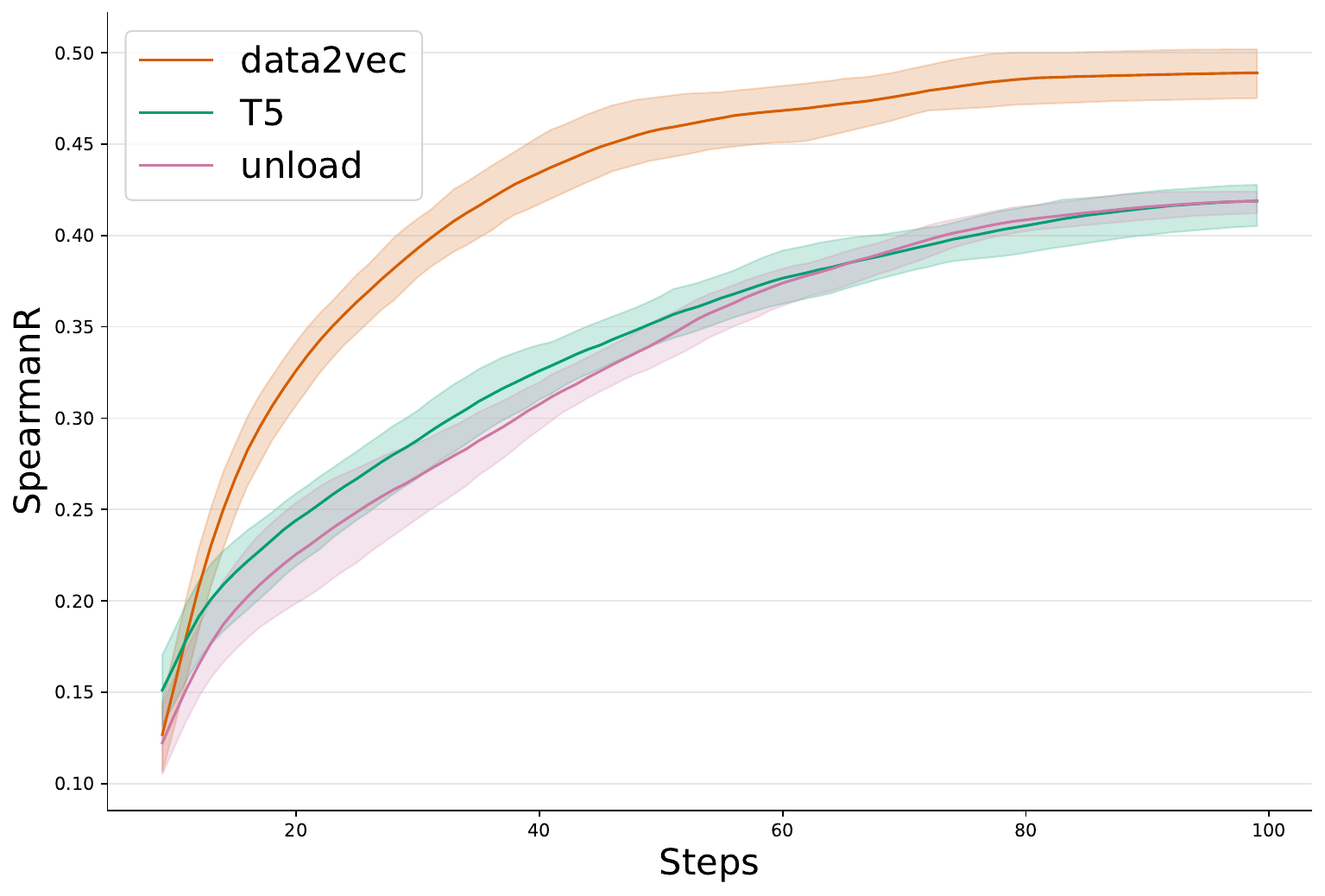}
         \caption{Translation efficiency (HEK) on 5$'$ UTR}
         \label{fig:hek_te_pretrain.5}
     \end{subfigure}
     \begin{subfigure}[t]{0.25\textwidth}
         \centering
         \includegraphics[width=\textwidth]{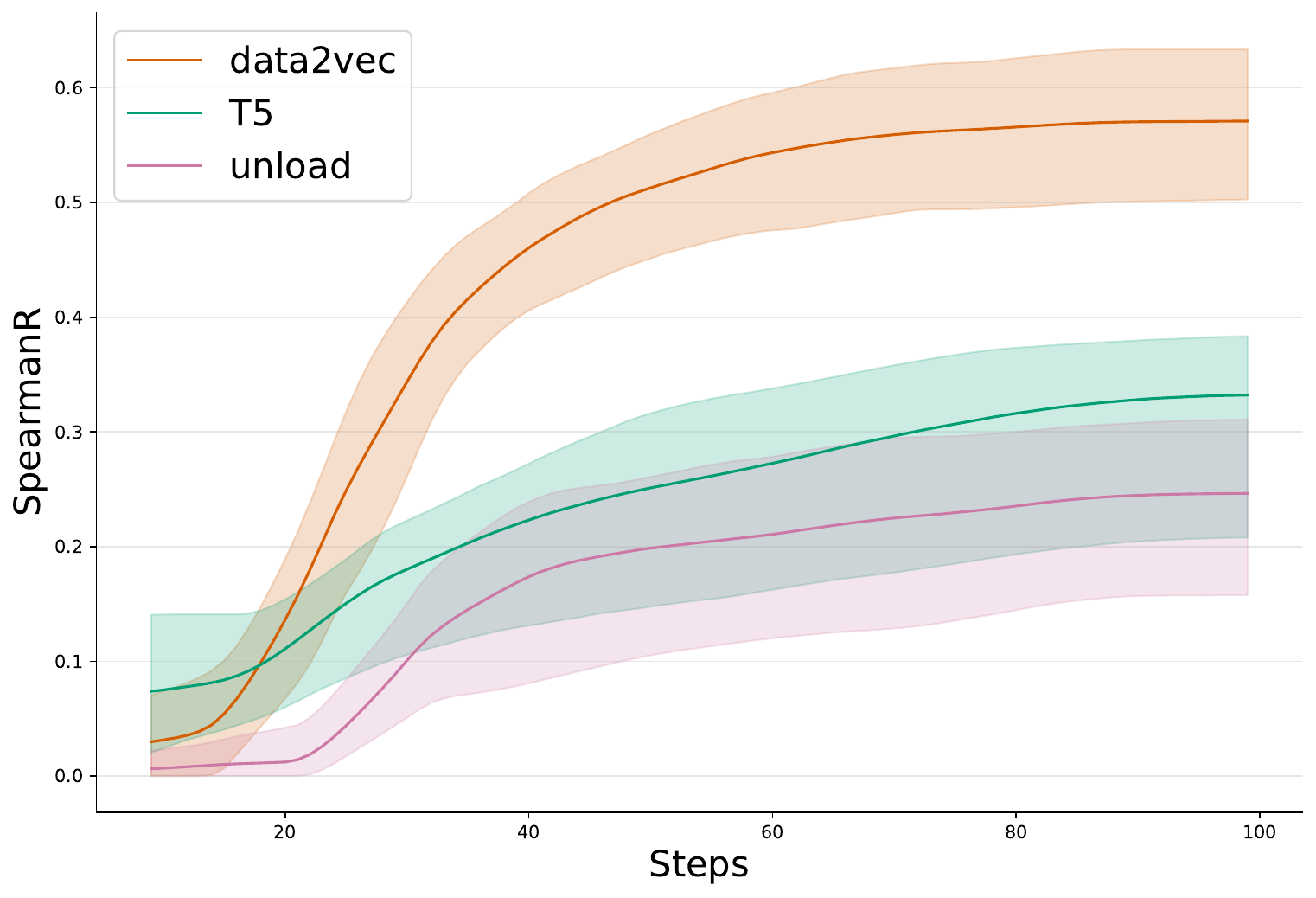}
         \caption{Expression level (Muscle) on 5$'$ UTR}
         \label{fig:muscle_el_pretrain}
     \end{subfigure}
     \begin{subfigure}[t]{0.25\textwidth}
         \centering
         \includegraphics[width=\textwidth]{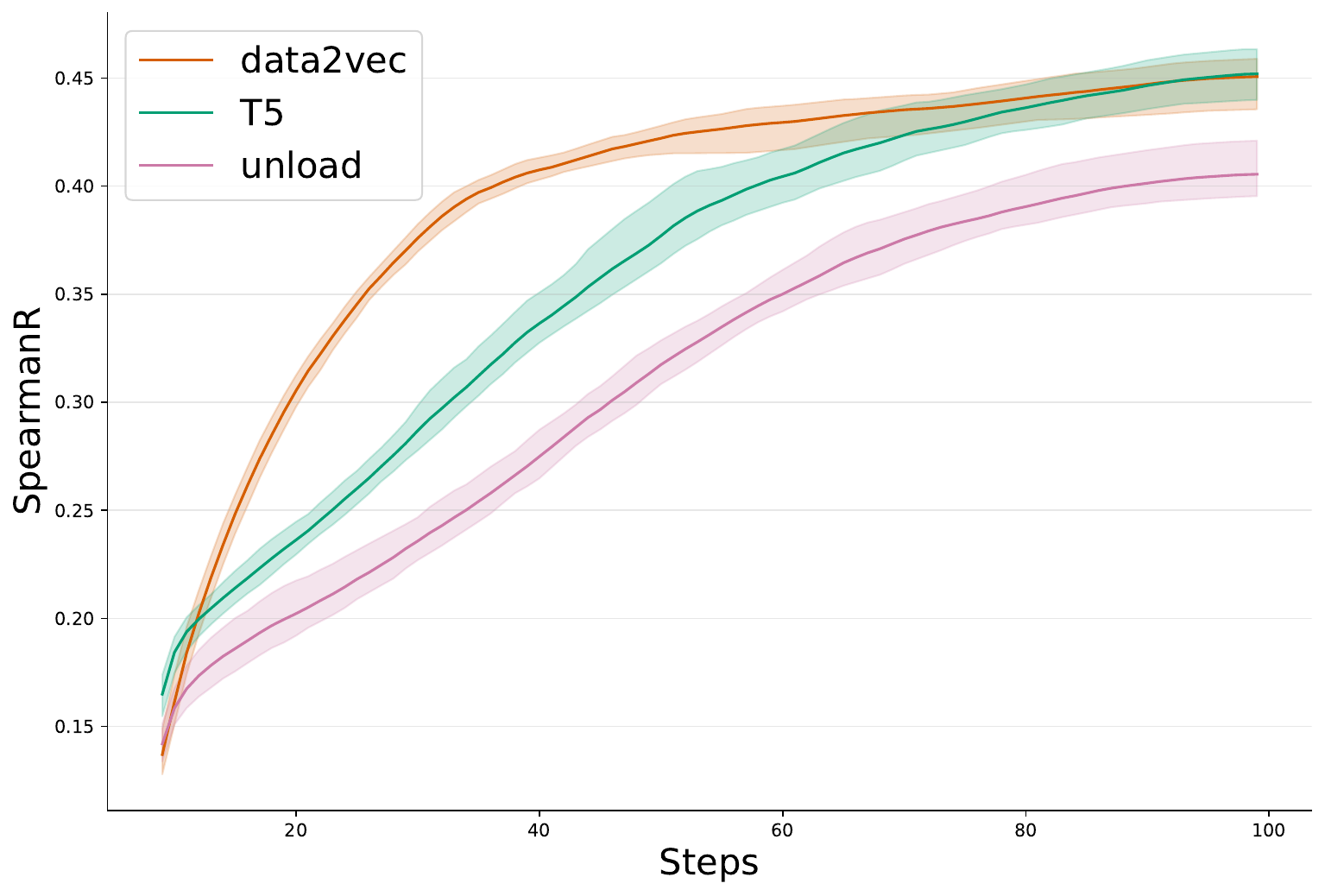}
         \caption{mRNA stability on CDS}
         \label{fig:sta_pretrain}
     \end{subfigure}
        \caption{Evaluate pretraining strategies:The data2vec with T5 encoder, the T5 encoder, and the untrained model (Unloaded)}
        \label{fig:pretrain_eva}
\end{figure*}
\subsection{Dataset}
For pretraining, we collect five species (human, rat, mouse, chicken, and zebrafish) mRNA sequences from the NIH with the datasets API \footnote{https://www.ncbi.nlm.nih.gov/datasets/docs/v2/reference-docs/command-line/datasets/}. 
To extract the 5$'$ UTR and CDS sequence from it, 
we discard the sequences after the stop codon (UAG, UAA, UGA). Thus, we have the 5$'$ UTR sequence before the start codon AUG and the CDS sequence after AUG until the stop codon. To avoid padding for the end of the area of 5$'$ UTR, we truncated the 5$'$ UTR sequence to a sequence with 102 nucleotides (34 tokens) from the start. After data processing, we obtain $510$k sequences with an average length of 459 bp. MFE value and SS are obtained using RNAfold from ViennaRNA\cite{lorenz2011viennarna}. 

For the downstream task on 5$'$ UTR data, we used three endogenous genes from three human cell types: Human embryonic kidney (HEK) 293T cell line, human prostate cancer cell line PC3 (PC3), and Human muscle tissue (Muscle) \cite{chu20245}. HEK and PC3 have over 10k sequences but the Muscle only has 1k dataset for the downstream task. The average length is 91 bp for 5 UTR data. 

For the downstream task CDS data, we used two datasets: the mRNA stability dataset \cite{diez2022icodon} and the mRFP Expression dataset \cite{nieuwkoop2023revealing}. The mRNA stability dataset contains 25K sequences for all endogenous genes in Humans, zebrafish, Xenopus, and mouse. The mRFP Expression dataset contains 6k sequences to measure protein production levels for several gene variants in E.coil.

\subsection{Experiment Setting}

The mRNA2vec pretraining architecture is built upon the data2vec framework, utilizing the T5 encoder as the student model. The model consists of 4 self-attention heads and 4 hidden layers, with each token represented by a 256-dimensional vector. In general,  our experiments were conducted using a model with 10M parameters. Since we used the Exponential Moving Average, approximately 6M parameters do not require gradients. The trainable parameters count is 3M. 
The predictor that maps the student model to the teacher model is a simple neural network composed of one linear layer, followed by a GELU activation layer, and another linear layer. The regressor $\phi$ is a three-layer neural network comprising one linear layer, one GELU activation layer, and a fully connected layer that maps the input to a scalar value. Similarly, the classifier $\psi$ is also a three-layer neural network, consisting of one linear layer, one GELU activation layer, and a fully connected layer, but it maps the input to a 27-dimensional output. We employed AdamW as the optimizer, with a learning rate of 0.001 and a weight decay of 0.001. The maximum sequence length was set to 128, and the batch size was 256. The pretraining runs on four Nvidia GeForce RTX 4090 GPUs.

\subsection{Pre-training Result}
\begin{table*}[h!]
\small
\centering
  \caption{Evaluate mRNA2vec pretrain strategies: The boldface is the best result of spearman rank value}
\begin{tabular}[h!]{ p{3.4cm}|p{1.cm} p{1cm}p{1cm}p{1cm}p{1cm}p{1cm}|p{2cm}p{2cm}}
 \hline
\multirow{2}*{Methods} & \multicolumn{6}{c|}{5$'$ UTR} & \multicolumn{2}{c}{CDS}\\
 \cline{2-9}
& TEK-TE &PC3-TE &Muscle-TE& TEK-EL &PC3-EL&Muscle-EL& mRNA-Stability& mRFP-Expression\\
 \hline
data2vec ($\beta_1 = \beta_2 = 0$)  &0.496	&0.540&0.550&0.490	&\textbf{0.488}&0.619&0.465&0.508	\\
data2vec+mfe ($\beta_2 = 0$)  &0.499	&0.559&0.572&0.518	&0.473&0.647&0.462	&0.549\\
data2vec+ss ($\beta_1 = 0$) &0.497	&0.554&0.551&0.491	&0.474&0.637&0.458&0.503\\
data2vec+mfe+ss   &\textbf{0.503}	&\textbf{0.560}&\textbf{0.573}&\textbf{0.521}	&\textbf{0.488}&\textbf{0.662}&\textbf{0.466}&\textbf{0.552}	\\
\hline
\end{tabular}
\label{tab:mRNA2vec pretrain}
\end{table*}
\subsubsection{Results}

In this section, we evaluate different pretraining strategies on mRNA sequences. After pretraining, the model weights are used to initialize the encoder. On top of the pretrained model, a downstream task regressor consisting of a linear layer, a GELU layer, and a fully connected layer is used to predict the target values. The encoder part is frozen, and only the parameters of the regressor are updated. The Spearman Rank is used to measure the regression result. 

First, we compare the data2vec approach using T5 as the encoder with a standalone T5 encoder on mRNA sequences. We also use an untrained (unloaded) model as the baseline, as shown in Figure \ref{fig:pretrain_eva}. Our observations are as follows:

\begin{enumerate}
    \item While the mRNA representation learned by T5 can be trained effectively in the first few epochs, the data2vec method shows better performance after several epochs on the 5$'$ UTR data. By incorporating the unmasked sequence during the pretraining process, data2vec extracts a more robust representation of the entire sequence, outperforming the results obtained by T5.
    
    \item Data2vec consistently improves results on both 5$'$ UTR and CDS regions compared to the untrained model baseline. However, T5 shows limited effectiveness on the 5$'$ UTR TE task. Contextual representation learning will have more advances for the mRNA compared to single mask token prediction such as T5.
\end{enumerate}

Next, we compare the performance of the pretrained model with and without the inclusion of MFE and SS-related losses. We evaluate four different pretraining models by adjusting \(\beta_1\) and \(\beta_2\). When \(\beta_1 = \beta_2 = 0\), the base model \(\mathcal{L}_{mRNA2vec} = \mathcal{L}\) does not incorporate MFE or SS information. When \(\beta_1 \ne 0\) and \(\beta_2 = 0\), the model \(\mathcal{L}_{mRNA2vec} = \mathcal{L} + \beta_1\mathcal{L}_{mfe}\) considers only MFE information. When \(\beta_1 = 0\) and \(\beta_2 \ne 0\), the model \(\mathcal{L}_{mRNA2vec} = \mathcal{L} + \beta_2\mathcal{L}_{ss}\) incorporates only SS information. 
As shown in Table \ref{tab:mRNA2vec pretrain}, 

\begin{enumerate}
    \item Incorporating both MFE and SS losses consistently improves the base model (without MFE and SS) across all downstream tasks. For instance, in the Muscle dataset, the TE task performance increased from 0.550 to 0.573, and the EL task performance improved from 0.619 to 0.663. This enhancement is attributed to the model's ability to consider sequence representation while aligning it with MFE and SS modalities.
    
    \item Analyzing the model under the scenarios where either \(\beta_1 = 0\) or \(\beta_2 = 0\), we observe that both cases generally improve the base model results. It is important to note that we used the same pretraining model for all tasks and datasets, with fixed hyperparameter combinations of \((\beta_1 = 0, \beta_2 = 0.001)\) and \((\beta_1 = 0.01, \beta_2 = 0)\). While this strategy may not represent the optimal hyperparameter settings for every dataset, it provides a more convenient and generalizable solution.
\end{enumerate}

\subsubsection{Discussion}
\begin{enumerate}
    \item \textbf{Choosing the optimal $\alpha$ for hard-masking}:
    Since certain sub-sequences of the mRNA sequence are more critical than others, we employ a specialized region masking strategy with the hyperparameter $\alpha$, as defined in Equation \ref{eq:q}. In this paper, we designate the important region spanning from the 15th token ($k=15$) to the 45th token ($k+h = 45$). To assess how $\alpha$ influences model learning, we experimented with different values ranging from 0 to 1, as shown in Table \ref{tab:mask_alpha}. When $\alpha$, the masking strategy is equivalent to random masking. However, when $\alpha>0$, the learning process accelerates, resulting in improved loss performance. It also enhances MFE learning, even without direct access to MFE information.
\begin{table}[h!]
\small
\centering
  \caption{Mask strategies}\small
\begin{tabular}[h!]{ p{1.1cm}ccccc  }
 \hline
$\alpha$&\textbf{0}& \textbf{0.2} &\textbf{0.5} &\textbf{0.8}&\textbf{1}\\
 \hline
$\mathcal{L}$&0.349   &0.348	&0.343&0.332&0.330	\\
SpermanR  &  -0.11	&-0.15&0.27&-0.09&-0.16	\\
\hline
\end{tabular}
\label{tab:mask_alpha}
\end{table}

    \item \textbf{Design a better pretext tasks}:
    Contrary to the findings of UTR-LM \cite{chu20245}, which suggested that incorporating secondary structure might degrade model performance, we explored a different approach by framing secondary structure as a token-wise classification problem with 27 categories (as opposed to the 3 categories used in UTR-LM). As shown in Figure \ref{fig:UTR-lmVSus}, the pretraining process benefits from the inclusion of MFE and secondary structure information when appropriate pretext tasks are employed.
\begin{figure}[h!]
     \centering
    \includegraphics[width=0.40\textwidth]{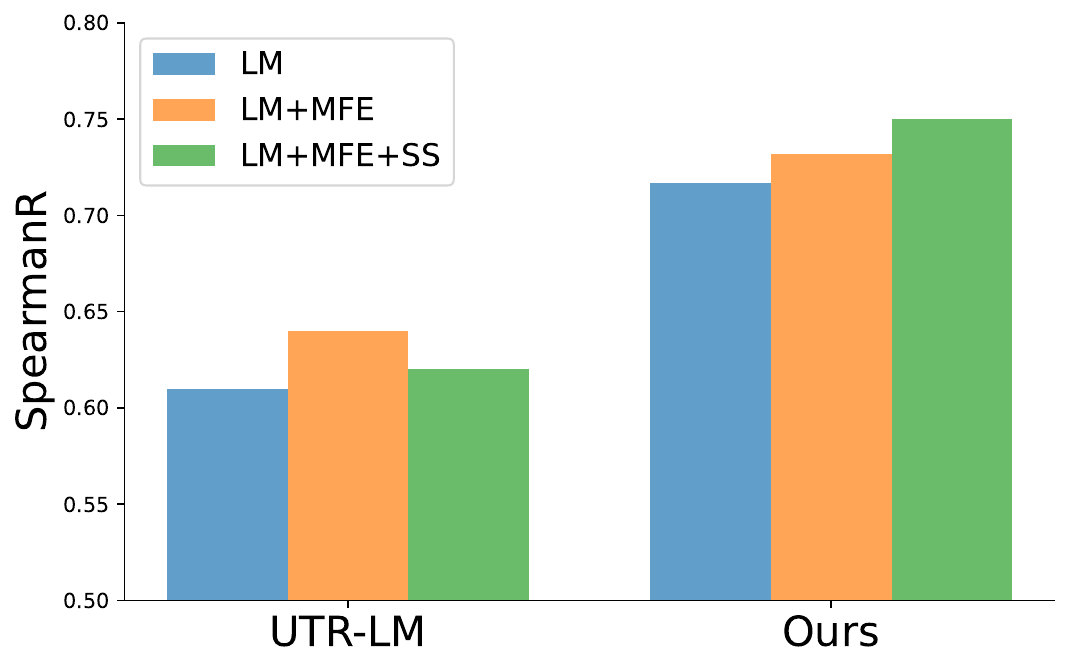}
         \caption{UTR-LM VS Ours}
         \label{fig:UTR-lmVSus}
\end{figure}
\end{enumerate}
\subsection{Downstream Tasks: Comparing with other methods}
\begin{table*}[t]
\small
\centering
  \caption{Experimental results on 5$'$ UTR. The boldface is the best result.}
\begin{tabular}[h!]{ p{4cm}p{1.5cm}p{1.5cm}p{1.5cm}p{1.5cm}p{1.5cm}p{1.5cm}  }
 \hline
 \textbf{Methods}& \textbf{HEK-TE} &\textbf{PC3-TE}&\textbf{Muscle-TE}&\textbf{HEK-EL}&\textbf{PC3-EL} & \textbf{Muscle-EL}\\
 \hline
RNA-FM \citet{chen2022interpretable} & 0.02	&0.0.3	&0.01	&0.13	&0.18	&0.05\\
RNABERT \citet{akiyama2022informative} & 0.40	&0.46	&0.56	&0.46	&0.44	&0.59\\
Cao-RF \citet{cao2021high} &0.55	&0.64	&0.64	&0.55	&0.57	&0.64\\
CodonBERT \citet{li2023codonbert}    & 0.56	&0.62	&0.64&0.56	&0.52	&0.64 \\
UTR-LM \citet{chu20245}  &  0.60	&0.63	&0.66	&0.65	&0.55	&0.61\\
RiNALMo\citet{penic2024rinalmo}  &  0.62	&0.70	&0.75	&0.58	&0.60	&0.78\\
mRNA2vec (Ours) & \textbf{0.68} & \textbf{0.71}& \textbf{0.75}	& \textbf{0.69}	& \textbf{0.70}	& \textbf{0.80}\\
 \hline
\end{tabular}
\label{tab:result}
\end{table*}
\begin{figure}[h!]
     \centering
    \includegraphics[width=0.40\textwidth]{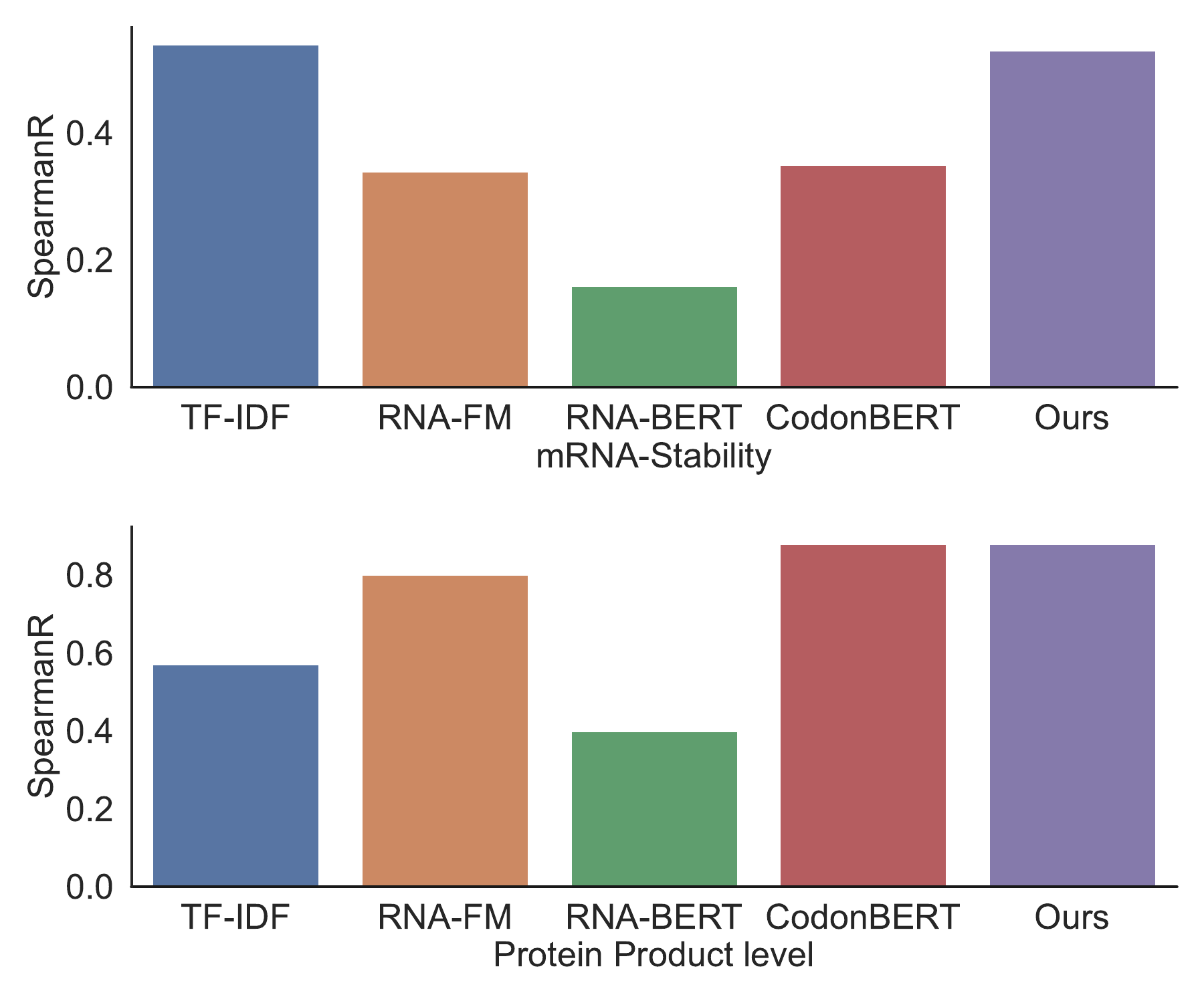}
         \caption{Compare our method with the methods working on CDS region}
         \label{fig:cds_result}
\end{figure}
\begin{figure*}[h!]
     \centering
     \begin{subfigure}[h!]{0.27\textwidth}
         \centering
         \includegraphics[width=\textwidth]{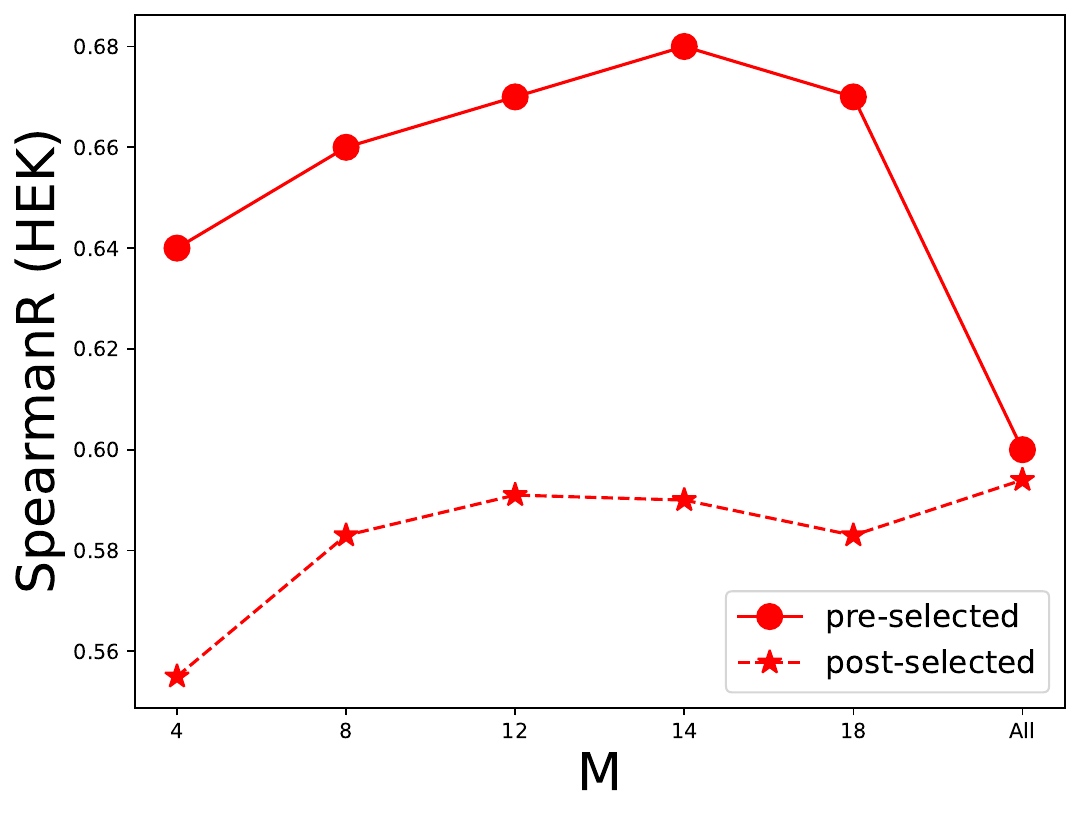}
         \label{fig:HEK_tokens}
     \end{subfigure}
     \begin{subfigure}[h!]{0.27\textwidth}
         \centering
         \includegraphics[width=\textwidth]{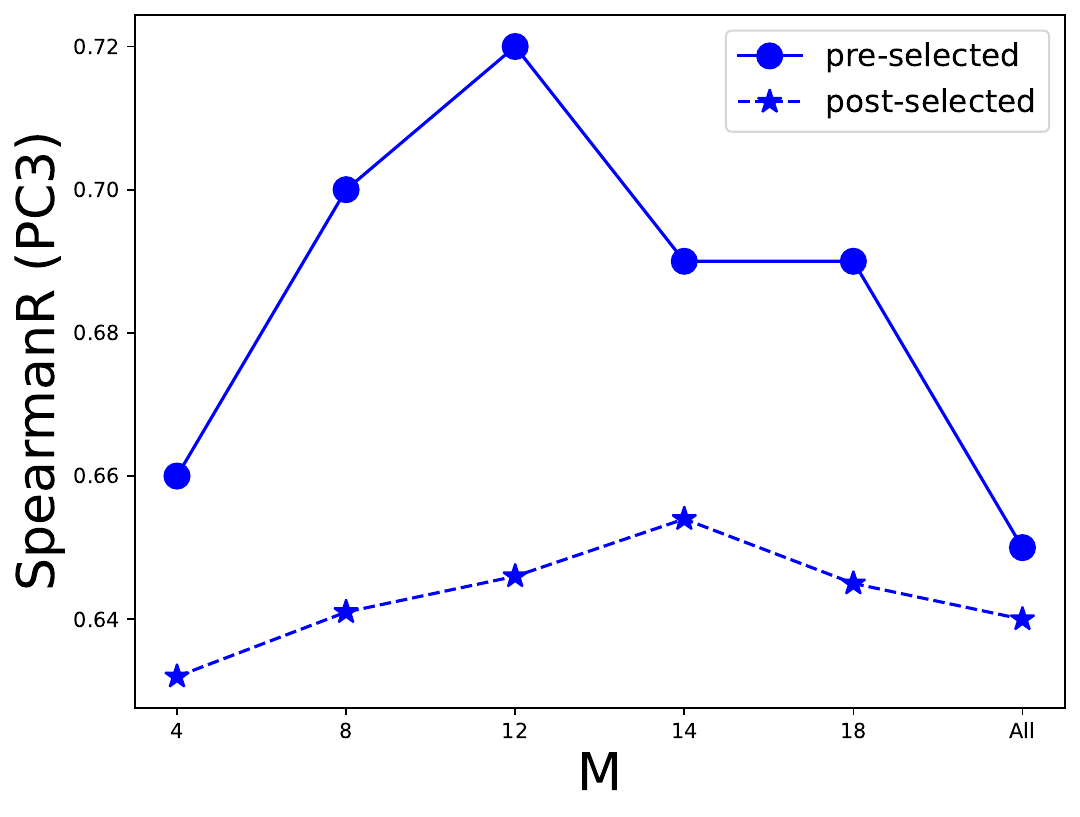}
         \label{fig:pc3_tokens}
     \end{subfigure}
     \begin{subfigure}[h!]{0.27\textwidth}
         \centering
         \includegraphics[width=\textwidth]{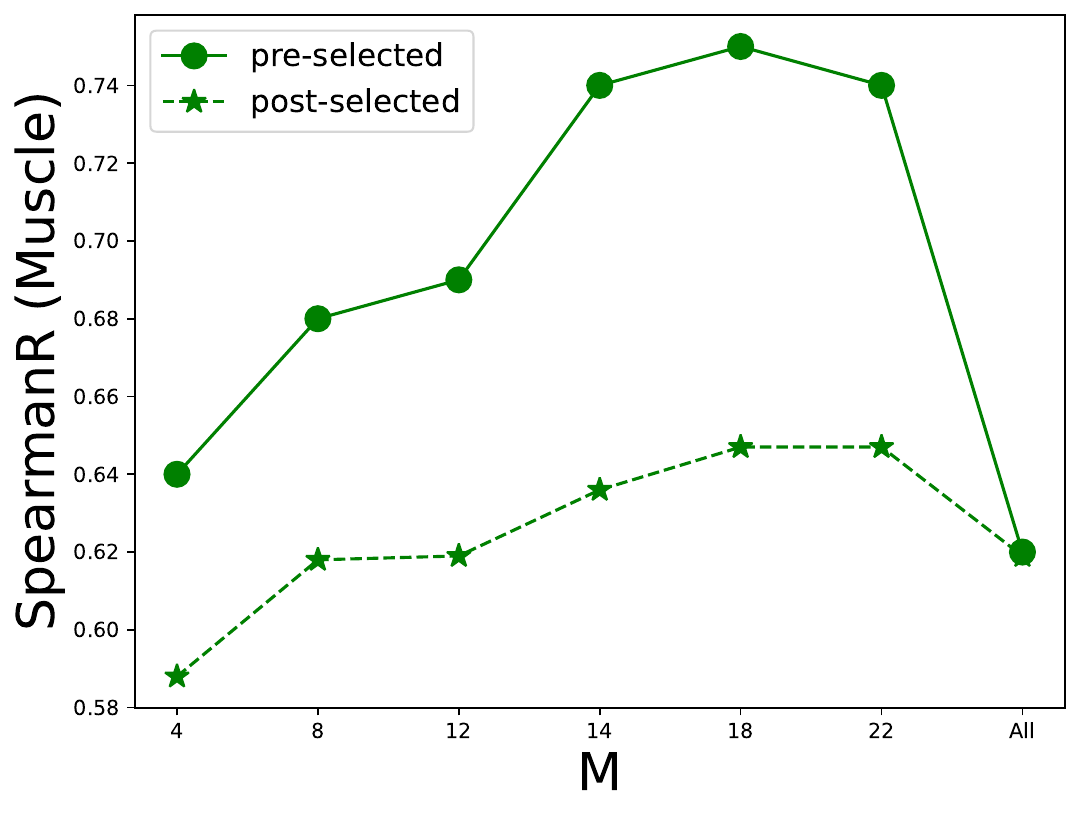}
         \label{fig:muscle_tokens}
     \end{subfigure}
        \caption{Two sub-sequence selection strategies. Pre-selected: change the input length of the encoder. Post-selected: change the output length of the encoder. The result will change as the sequence length changes.}
        \label{fig:diff_tokens}
\end{figure*}
In this section, we compare our method to several benchmarks that focus on TE and EL prediction by measuring the Spearman rank for the 5$'$ UTR:
1) CodonBERT \cite{li2023codonbert}, 
2) UTR-LM \cite{chu20245}, 
3) RNA-BERT \cite{akiyama2022informative}, 
4) RNA-FM \cite{chen2022interpretable}, 
5) Cao-RF \cite{cao2021high},
6) RiNALMo \cite{penic2024rinalmo}.

We also compare our method to existing approaches that work on function of sequence for the CDS:
1) Term Frequency–Inverse Document Frequency (TF-IDF) \cite{rajaraman2011mining}, 
2) RNA-BERT \cite{akiyama2022informative}, 
3) RNA-FM \cite{chen2022interpretable}, 
4) CodonBERT \cite{li2023codonbert}.

\subsubsection{Results}

For the 5$'$ UTR dataset, after obtaining the pre-trained model, we fine-tuned it on three 5$'$ UTR datasets. As shown in Table \ref{tab:result}, our method outperforms all other approaches on datasets related to the TE and EL tasks. Compared to the same size of model UTR-LM on TE task,  we achieved improvements of 13\%, 12\%, and 14\%  on the HEK, PC3, and Muscle datasets, respectively. For the EL task, our method improvements of 6\%, 27\%, and 31\%.

For the CDS dataset, we evaluate these methods on the mRNA stability dataset and the mRFP-Expression dataset (protein production level of \textit{E. coli}). As shown in Figure \ref{fig:cds_result}, our method significantly improves the results compared to the codon representation method CodonBERT from 0.34 to 0.53. For protein production levels, our method achieves comparable results. However, unlike CodonBERT, our pretraining did not use \textit{E. coli} mRNA sequences. This indicates that our pre-trained model can be applied to species beyond the five species used in the pretraining.



\subsubsection{Discussion}
With our method, the SOTA results for TE and EL tasks have been significantly improved. Our method also achieved comparable results in the CDS region. We have four key observations that contribute to the performance of the downstream tasks.

\begin{enumerate}
    \item \textbf{Finding an optimal input sequence length:}
    We examined the effect of different input sequence lengths. The 5$'$ UTR datasets (HEK, PC3, Muscle) have an average length of 91 nucleotides (31 tokens after tokenization). We truncated the sequence from the left and kept only the rightmost $M$ tokens. The solid line in Figure \ref{fig:diff_tokens} shows the TE results on the different datasets according to $M$. Input sequence length affects the TE outcome, with full sequences worsening the results. This finding supports the idea that only a few bases in the 5$'$ UTR play a crucial role in translation \cite{nieuwkoop2023revealing}.

    Instead of selecting sub-sequences by truncating the input sequence, we directly selected the representation of the sub-sequence by choosing the tokens after pretraining, as shown by the dashed line in Figure \ref{fig:diff_tokens}. Both curves suggest that 12, 14, and 18 tokens might be the optimal input sequence lengths for HEK, PC3, and Muscle, respectively. However, this approach does not perform as well as the strategy of controlling the input length.

    \item \textbf{Selecting hidden state:}
    Different hidden layers provide other representations of the mRNA sequence \cite{chen2020simple}. We studied the impact of selecting the hidden layer used for the downstream tasks. In Table \ref{tab:hidden_state}, using the last hidden state ($K=-1$) harms the performance compared to using the second last hidden layer ($K = -2$) or the third last hidden state($K = -3$). This paper uses the second last hidden state as the mRNA embedding.

    \item \textbf{Designing a better downstream task regressor:}
    We compare different regression heads on top of the pre-trained model, including a linear regressor, a simple two-layer neural network, and a convolutional neural network. As shown in Table \ref{tab:head}, the linear regressor can outperform the UTR-LM results in Table \ref{tab:result}. While more complex regressors generally improve performance over linear regressors, their advantage diminishes with larger datasets, such as HEK.

\begin{table}[h!]
\small
\centering
  \caption{Different embedding layers}
\begin{tabular}[h!]{ p{1cm}ccc  }
 \hline
 $K$& \textbf{HEK} &\textbf{PC3} &\textbf{Muscle}\\
 \hline
-1    &0.389	&0.556&0.560	\\
-2  &  \textbf{0.515}	&\textbf{0.619} & \textbf{0.66}	\\
-3  &  0.504	&0.598&0.640	\\
\hline
\end{tabular}
\label{tab:hidden_state}
\end{table}
\begin{table}[h!]
\small
\centering
  \caption{Regressor Head comparison on TE}
\begin{tabular}[h!]{ p{2cm}ccc  }
 \hline
 \textbf{Regressor}& \textbf{HEK} &\textbf{PC3} &\textbf{Muscle}\\
 \hline
linear    &0.668	&0.692&0.715	\\
Two-Layer  & 0.669	&0.713&0.752	\\
CNN  &  0.682	&0.709&0.761	\\
\hline
\end{tabular}
\label{tab:head}
\end{table}
\end{enumerate}
\section{Conclusion}
In this paper, we proposed a novel mRNA pretraining method, mRNA2vec, based on the data2vec architecture and mRNA domain knowledge. Due to the contextual learning and hard-mask strategy, the data2vec-based LM outperforms the mask token prediction LMs such as T5. We also find that using the domain knowledge MFE and SS properly can improve the learning result. Meanwhile, with optimal sub-sequence, better-hidden layer, and a well-designed regressor for the downstream task, mRNA2vec improves the best benchmark largely on the TE and EL task and obtains comparable results on the 
mRNA stability and \textit{E.coil} protein production. 

\section{Acknowledgments}
This work was carried out during an internship at XtaiPi Inc.
This work was supported by the National Science Foundation AI Institute in Dynamic Systems (Grant No. 2112085).

\bibliography{aaai25}

\begin{thebibliography}{30}
\providecommand{\natexlab}[1]{#1}

\bibitem[{Akiyama and Sakakibara(2022)}]{akiyama2022informative}
Akiyama, M.; and Sakakibara, Y. 2022.
\newblock Informative RNA base embedding for RNA structural alignment and clustering by deep representation learning.
\newblock \emph{NAR genomics and bioinformatics}, 4(1): lqac012.

\bibitem[{Baevski et~al.(2022)Baevski, Hsu, Xu, Babu, Gu, and Auli}]{baevski2022data2vec}
Baevski, A.; Hsu, W.-N.; Xu, Q.; Babu, A.; Gu, J.; and Auli, M. 2022.
\newblock Data2vec: A general framework for self-supervised learning in speech, vision and language.
\newblock In \emph{International Conference on Machine Learning}, 1298--1312. PMLR.

\bibitem[{Cao et~al.(2021)Cao, Novoa, Zhang, Chen, Liu, Choi, Wong, Wehrspaun, Kellis, and Lu}]{cao2021high}
Cao, J.; Novoa, E.~M.; Zhang, Z.; Chen, W.~C.; Liu, D.; Choi, G.~C.; Wong, A.~S.; Wehrspaun, C.; Kellis, M.; and Lu, T.~K. 2021.
\newblock High-throughput 5$'$UTR engineering for enhanced protein production in non-viral gene therapies.
\newblock \emph{Nature communications}, 12(1): 4138.

\bibitem[{Cetnar and Salis(2021)}]{2021Systematic}
Cetnar, D.~P.; and Salis, H.~M. 2021.
\newblock Systematic Quantification of Sequence and Structural Determinants Controlling mRNA stability in Bacterial Operons.
\newblock \emph{ACS Synthetic Biology}, 10(2): 318--332.

\bibitem[{Chen et~al.(2022)Chen, Hu, Sun, Tan, Wang, Yu, Zong, Hong, Xiao, Shen et~al.}]{chen2022interpretable}
Chen, J.; Hu, Z.; Sun, S.; Tan, Q.; Wang, Y.; Yu, Q.; Zong, L.; Hong, L.; Xiao, J.; Shen, T.; et~al. 2022.
\newblock Interpretable RNA foundation model from unannotated data for highly accurate RNA structure and function predictions.
\newblock \emph{arXiv preprint arXiv:2204.00300}.

\bibitem[{Chen et~al.(2020)Chen, Kornblith, Norouzi, and Hinton}]{chen2020simple}
Chen, T.; Kornblith, S.; Norouzi, M.; and Hinton, G. 2020.
\newblock A simple framework for contrastive learning of visual representations.
\newblock In \emph{International conference on machine learning}, 1597--1607. PMLR.

\bibitem[{Chu et~al.(2024)Chu, Yu, Li, Huang, Shen, Cong, Zhang, and Wang}]{chu20245}
Chu, Y.; Yu, D.; Li, Y.; Huang, K.; Shen, Y.; Cong, L.; Zhang, J.; and Wang, M. 2024.
\newblock A 5$'$UTR language model for decoding untranslated regions of mRNA and function predictions.
\newblock \emph{Nature Machine Intelligence}, 6(4): 449--460.

\bibitem[{Devlin et~al.(2018)Devlin, Chang, Lee, and Toutanova}]{devlin2018bert}
Devlin, J.; Chang, M.-W.; Lee, K.; and Toutanova, K. 2018.
\newblock Bert: Pre-training of deep bidirectional transformers for language understanding.
\newblock \emph{arXiv preprint arXiv:1810.04805}.

\bibitem[{Diez et~al.(2022)Diez, Medina-Mu{\~n}oz, Castellano, da~Silva~Pescador, Wu, and Bazzini}]{diez2022icodon}
Diez, M.; Medina-Mu{\~n}oz, S.~G.; Castellano, L.~A.; da~Silva~Pescador, G.; Wu, Q.; and Bazzini, A.~A. 2022.
\newblock iCodon customizes gene expression based on the codon composition.
\newblock \emph{Scientific Reports}, 12(1): 12126.

\bibitem[{He et~al.(2020)He, Fan, Wu, Xie, and Girshick}]{he2020momentum}
He, K.; Fan, H.; Wu, Y.; Xie, S.; and Girshick, R. 2020.
\newblock Momentum contrast for unsupervised visual representation learning.
\newblock In \emph{Proceedings of the IEEE/CVF conference on computer vision and pattern recognition}, 9729--9738.

\bibitem[{Huang et~al.(2023)Huang, Ferber, Tian, Dilkina, and Steiner}]{huang2023searching}
Huang, T.; Ferber, A.~M.; Tian, Y.; Dilkina, B.; and Steiner, B. 2023.
\newblock Searching large neighborhoods for integer linear programs with contrastive learning.
\newblock In \emph{International Conference on Machine Learning}, 13869--13890. PMLR.

\bibitem[{Jackson et~al.(2020)Jackson, Anderson, Rouphael, Roberts, Makhene, Coler, McCullough, Chappell, Denison, Stevens et~al.}]{jackson2020mrna}
Jackson, L.~A.; Anderson, E.~J.; Rouphael, N.~G.; Roberts, P.~C.; Makhene, M.; Coler, R.~N.; McCullough, M.~P.; Chappell, J.~D.; Denison, M.~R.; Stevens, L.~J.; et~al. 2020.
\newblock An mRNA vaccine against SARS-CoV-2—preliminary report.
\newblock \emph{New England journal of medicine}, 383(20): 1920--1931.

\bibitem[{Kwon et~al.(2018)Kwon, Kim, Seo, Moon, Lee, Lee, and Lee}]{kwon2018emergence}
Kwon, H.; Kim, M.; Seo, Y.; Moon, Y.~S.; Lee, H.~J.; Lee, K.; and Lee, H. 2018.
\newblock Emergence of synthetic mRNA: In vitro synthesis of mRNA and its applications in regenerative medicine.
\newblock \emph{Biomaterials}, 156: 172--193.

\bibitem[{Leppek et~al.(2022)Leppek, Byeon, Kladwang, Wayment-Steele, Kerr, Xu, Kim, Topkar, Choe, Rothschild et~al.}]{leppek2022combinatorial}
Leppek, K.; Byeon, G.~W.; Kladwang, W.; Wayment-Steele, H.~K.; Kerr, C.~H.; Xu, A.~F.; Kim, D.~S.; Topkar, V.~V.; Choe, C.; Rothschild, D.; et~al. 2022.
\newblock Combinatorial optimization of mRNA structure, stability, and translation for RNA-based therapeutics.
\newblock \emph{Nature communications}, 13(1): 1536.

\bibitem[{Li et~al.(2023)Li, Moayedpour, Li, Bailey, Riahi, Kogler-Anele, Miladi, Miner, Zheng, Wang et~al.}]{li2023codonbert}
Li, S.; Moayedpour, S.; Li, R.; Bailey, M.; Riahi, S.; Kogler-Anele, L.; Miladi, M.; Miner, J.; Zheng, D.; Wang, J.; et~al. 2023.
\newblock Codonbert: Large language models for mrna design and optimization.
\newblock \emph{bioRxiv}, 2023--09.

\bibitem[{Lorenz et~al.(2011)Lorenz, Bernhart, H{\"o}ner~zu Siederdissen, Tafer, Flamm, Stadler, and Hofacker}]{lorenz2011viennarna}
Lorenz, R.; Bernhart, S.~H.; H{\"o}ner~zu Siederdissen, C.; Tafer, H.; Flamm, C.; Stadler, P.~F.; and Hofacker, I.~L. 2011.
\newblock ViennaRNA Package 2.0.
\newblock \emph{Algorithms for molecular biology}, 6: 1--14.

\bibitem[{Mikolov et~al.(2013)Mikolov, Chen, Corrado, and Dean}]{mikolov2013efficient}
Mikolov, T.; Chen, K.; Corrado, G.; and Dean, J. 2013.
\newblock Efficient estimation of word representations in vector space.
\newblock \emph{arXiv preprint arXiv:1301.3781}.

\bibitem[{Ng(2017)}]{ng2017dna2vec}
Ng, P. 2017.
\newblock dna2vec: Consistent vector representations of variable-length k-mers.
\newblock \emph{arXiv preprint arXiv:1701.06279}.

\bibitem[{Nieuwkoop et~al.(2023)Nieuwkoop, Terlouw, Stevens, Scheltema, De~Ridder, Van~der Oost, and Claassens}]{nieuwkoop2023revealing}
Nieuwkoop, T.; Terlouw, B.~R.; Stevens, K.~G.; Scheltema, R.~A.; De~Ridder, D.; Van~der Oost, J.; and Claassens, N.~J. 2023.
\newblock Revealing determinants of translation efficiency via whole-gene codon randomization and machine learning.
\newblock \emph{Nucleic acids research}, 51(5): 2363--2376.

\bibitem[{Nikolados and Oyarzun(2023)}]{2023Deep}
Nikolados, E.~M.; and Oyarzun, D.~A. 2023.
\newblock Deep learning for optimization of protein expression.
\newblock \emph{Current opinion in biotechnology}.

\bibitem[{Oord, Li, and Vinyals(2018)}]{oord2018representation}
Oord, A. v.~d.; Li, Y.; and Vinyals, O. 2018.
\newblock Representation learning with contrastive predictive coding.
\newblock \emph{arXiv preprint arXiv:1807.03748}.

\bibitem[{Peni{\'c} et~al.(2024)Peni{\'c}, Vla{\v{s}}i{\'c}, Huber, Wan, and {\v{S}}iki{\'c}}]{penic2024rinalmo}
Peni{\'c}, R.~J.; Vla{\v{s}}i{\'c}, T.; Huber, R.~G.; Wan, Y.; and {\v{S}}iki{\'c}, M. 2024.
\newblock Rinalmo: General-purpose rna language models can generalize well on structure prediction tasks.
\newblock \emph{arXiv preprint arXiv:2403.00043}.

\bibitem[{Raffel et~al.(2020)Raffel, Shazeer, Roberts, Lee, Narang, Matena, Zhou, Li, and Liu}]{raffel2020exploring}
Raffel, C.; Shazeer, N.; Roberts, A.; Lee, K.; Narang, S.; Matena, M.; Zhou, Y.; Li, W.; and Liu, P.~J. 2020.
\newblock Exploring the limits of transfer learning with a unified text-to-text transformer.
\newblock \emph{Journal of machine learning research}, 21(140): 1--67.

\bibitem[{Rajaraman and Ullman(2011)}]{rajaraman2011mining}
Rajaraman, A.; and Ullman, J.~D. 2011.
\newblock \emph{Mining of massive datasets}.
\newblock Autoedicion.

\bibitem[{Rao et~al.(2020)Rao, Meier, Sercu, Ovchinnikov, and Rives}]{rao2020transformer}
Rao, R.; Meier, J.; Sercu, T.; Ovchinnikov, S.; and Rives, A. 2020.
\newblock Transformer protein language models are unsupervised structure learners.
\newblock \emph{Biorxiv}, 2020--12.

\bibitem[{Rubachev et~al.(2022)Rubachev, Alekberov, Gorishniy, and Babenko}]{rubachev2022revisiting}
Rubachev, I.; Alekberov, A.; Gorishniy, Y.; and Babenko, A. 2022.
\newblock Revisiting pretraining objectives for tabular deep learning.
\newblock \emph{arXiv preprint arXiv:2207.03208}.

\bibitem[{Sample et~al.(2018)Sample, Wang, Reid, Presnyak, and Seelig}]{2018Human}
Sample, P.~J.; Wang, B.; Reid, D.~W.; Presnyak, V.; and Seelig, G. 2018.
\newblock Human 5$'$UTR design and variant effect prediction from a massively parallel translation assay.
\newblock \emph{Cold Spring Harbor Laboratory}, (7).

\bibitem[{To and Cho(2021)}]{to2021overview}
To, K.~K.; and Cho, W.~C. 2021.
\newblock An overview of rational design of mRNA-based therapeutics and vaccines.
\newblock \emph{Expert opinion on drug discovery}, 16(11): 1307--1317.

\bibitem[{Wieder et~al.(2024)Wieder, D’Souza, Martin-Geary, Lassen, Talbot-Martin, Fernandes, Chothani, Rackham, Schafer, Aspden et~al.}]{wieder2024differences}
Wieder, N.; D’Souza, E.~N.; Martin-Geary, A.~C.; Lassen, F.~H.; Talbot-Martin, J.; Fernandes, M.; Chothani, S.~P.; Rackham, O.~J.; Schafer, S.; Aspden, J.~L.; et~al. 2024.
\newblock Differences in 5'untranslated regions highlight the importance of translational regulation of dosage sensitive genes.
\newblock \emph{Genome Biology}, 25(1): 111.

\bibitem[{Zeraati et~al.(2017)Zeraati, Moye, Wong, Perera, Cowley, Christ, Bryan, and Dinger}]{zeraati2017cancer}
Zeraati, M.; Moye, A.~L.; Wong, J.~W.; Perera, D.; Cowley, M.~J.; Christ, D.~U.; Bryan, T.~M.; and Dinger, M.~E. 2017.
\newblock Cancer-associated noncoding mutations affect RNA G-quadruplex-mediated regulation of gene expression.
\newblock \emph{Scientific reports}, 7(1): 708.

\end{thebibliography}
\section*{Downstream Task Dataset}
As shown in Table \ref{tab:dataset}, we have 3 5$'$ UTR downstream task datasets and 2 CDS downstream task datasets. The HEK and PC3 are larger than the Muscle. And the mRNA-stability is larger than mRFP-Expression. 
\begin{table}[h!]
\centering
  \caption{Dataset detail.}
\begin{tabular}[h!]{ p{0.6cm}ccccc  }
 \hline
 &\textbf{HEK}& \textbf{PC3} &\textbf{Muscle} &\textbf{Stability}&\textbf{mRFP-Exp}\\
 \hline
\#seqs   	&14k&12k&1k&25k&1.5k\\
 \hline
Average-length  &91&91&91&670&  678		\\
\hline
\end{tabular}
\label{tab:dataset}
\end{table}

\section*{SS Boosts Small Dataset More}
\begin{figure}[t]
     \centering
    \includegraphics[width=0.40\textwidth]{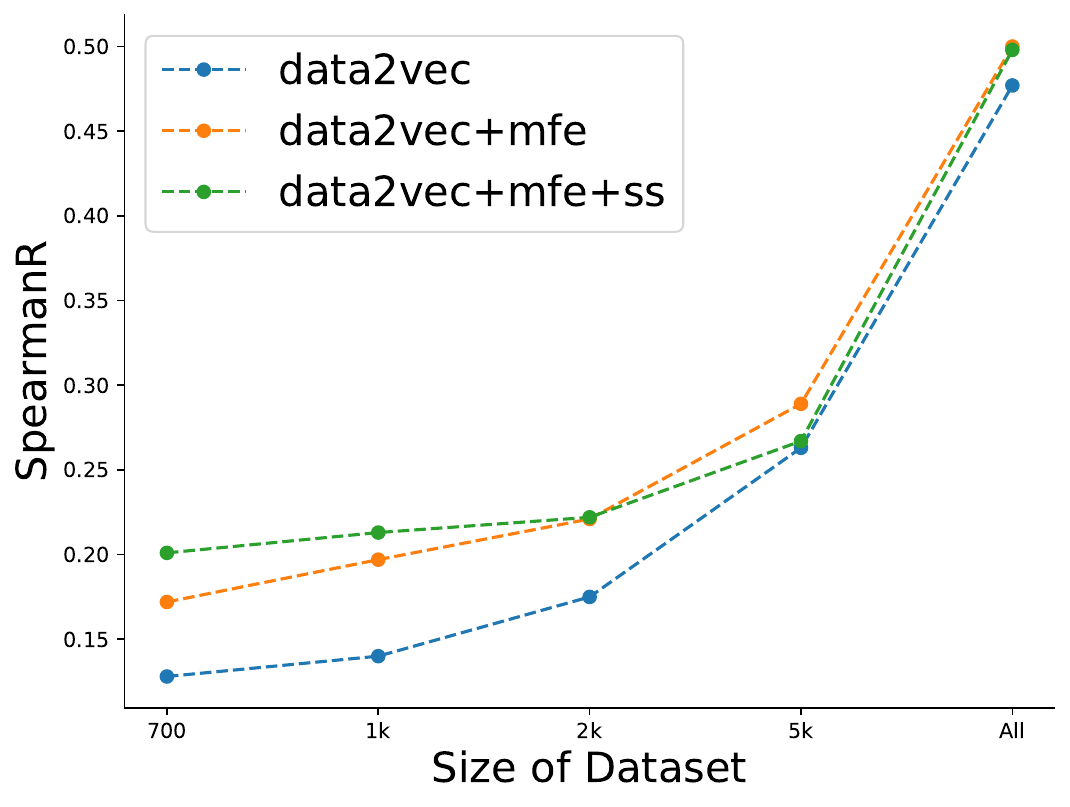}
    \caption{Evaluate different pretrain strategies on 5$'$ UTR region}
    \label{fig:small data}
\end{figure}

We have observed in Table \ref{tab:mRNA2vec pretrain} that MFE and SS can better improve the performance on Muscle data than other datasets. Thus, we hypothesized that MFE and SS can contribute more on the small dataset. We downsampled the HEK dataset to train different regressors with the same setting of the above experiments.  Figure \ref{fig:small data} shows the Spearman rank of TE on the test dataset after using a different sub-dataset as the training dataset. Our hypothesis could be verified by comparing the based model+MFE loss, model+MFE loss+ SS loss, and the base model. The margins between the based model and other models will be increased as the training data becomes smaller. 

\section*{More results of Evaluation of Downstream Task Regressor}
As shown in Table \ref{tab:head_el}, it is consistent with Table \ref{tab:head}. More complex regressors generally improve performance over linear regressors.
\begin{table}[t]
\small
\centering
  \caption{Regressor Head comparison on EL}
\begin{tabular}[t]{ p{2cm}p{1.5cm}p{1.5cm}p{1.5cm}  }
 \hline
 \textbf{Regressor}& \textbf{HEK} &\textbf{PC3} &\textbf{Muscle}\\
 \hline
linear    &0.683	&0.698&0.780	\\
Two-layer  & 0.685	&0.700&0.750	\\
CNN  &  0.693	&0.702&0.798	\\
\hline
\end{tabular}
\label{tab:head_el}
\end{table}



\end{document}